\documentclass[12pt]{article}
\usepackage{html}
\usepackage{epsf}
\usepackage{graphicx}
\usepackage{amssymb}
\newcommand{\Msun}{\ensuremath{M_{\mathrm{Sun}}}}
\newcommand{\densc}{\ensuremath{\rho_{\mathrm{c}}}}
\newcommand{\densmax}{\ensuremath{\rho_{\mathrm{c,max}}}}
\newcommand{\Mmax}{\ensuremath{M_{\mathrm{max}}}}
\begin{document}
\title{MATTERS OF GRAVITY, The newsletter of the APS Topical Group on 
Gravitation}
\begin{center}
{ \Large {\bf MATTERS OF GRAVITY}}\\ 
\bigskip
\hrule
\medskip
{The newsletter of the Topical Group on Gravitation of the American Physical 
Society}\\
\medskip
{\bf Number 38 \hfill Fall 2011}
\end{center}
\begin{flushleft}
\tableofcontents
\vfill\eject
\section*{\noindent  Editor\hfill}
David Garfinkle\\
\smallskip
Department of Physics
Oakland University
Rochester, MI 48309\\
Phone: (248) 370-3411\\
Internet: 
\htmladdnormallink{\protect {\tt{garfinkl-at-oakland.edu}}}
{mailto:garfinkl@oakland.edu}\\
WWW: \htmladdnormallink
{\protect {\tt{http://www.oakland.edu/?id=10223\&sid=249\#garfinkle}}}
{http://www.oakland.edu/?id=10223&sid=249\#garfinkle}\\

\section*{\noindent  Associate Editor\hfill}
Greg Comer\\
\smallskip
Department of Physics and Center for Fluids at All Scales,\\
St. Louis University,
St. Louis, MO 63103\\
Phone: (314) 977-8432\\
Internet:
\htmladdnormallink{\protect {\tt{comergl-at-slu.edu}}}
{mailto:comergl@slu.edu}\\
WWW: \htmladdnormallink{\protect {\tt{http://www.slu.edu/colleges/AS/physics/profs/comer.html}}}
{http://www.slu.edu//colleges/AS/physics/profs/comer.html}\\
\bigskip
\hfill ISSN: 1527-3431

\bigskip

DISCLAIMER: The opinions expressed in the articles of this newsletter represent
the views of the authors and are not necessarily the views of APS.
The articles in this newsletter are not peer reviewed.

\begin{rawhtml}
<P>
<BR><HR><P>
\end{rawhtml}
\end{flushleft}
\pagebreak
\section*{Editorial}

The next newsletter is due February 1st.  This and all subsequent
issues will be available on the web at
\htmladdnormallink 
{\protect {\tt {https://files.oakland.edu/users/garfinkl/web/mog/}}}
{https://files.oakland.edu/users/garfinkl/web/mog/} 
All issues before number {\bf 28} are available at
\htmladdnormallink {\protect {\tt {http://www.phys.lsu.edu/mog}}}
{http://www.phys.lsu.edu/mog}

Any ideas for topics
that should be covered by the newsletter, should be emailed to me, or 
Greg Comer, or
the relevant correspondent.  Any comments/questions/complaints
about the newsletter should be emailed to me.

A hardcopy of the newsletter is distributed free of charge to the
members of the APS Topical Group on Gravitation upon request (the
default distribution form is via the web) to the secretary of the
Topical Group.  It is considered a lack of etiquette to ask me to mail
you hard copies of the newsletter unless you have exhausted all your
resources to get your copy otherwise.

\hfill David Garfinkle 

\bigbreak

\vspace{-0.8cm}
\parskip=0pt
\section*{Correspondents of Matters of Gravity}
\begin{itemize}
\setlength{\itemsep}{-5pt}
\setlength{\parsep}{0pt}
\item John Friedman and Kip Thorne: Relativistic Astrophysics,
\item Bei-Lok Hu: Quantum Cosmology and Related Topics
\item Veronika Hubeny: String Theory
\item Beverly Berger: News from NSF
\item Luis Lehner: Numerical Relativity
\item Jim Isenberg: Mathematical Relativity
\item Katherine Freese: Cosmology
\item Lee Smolin: Quantum Gravity
\item Cliff Will: Confrontation of Theory with Experiment
\item Peter Bender: Space Experiments
\item Jens Gundlach: Laboratory Experiments
\item Warren Johnson: Resonant Mass Gravitational Wave Detectors
\item David Shoemaker: LIGO Project
\item Stan Whitcomb: Gravitational Wave detection
\item Peter Saulson and Jorge Pullin: former editors, correspondents at large.
\end{itemize}
\section*{Topical Group in Gravitation (GGR) Authorities}
Chair: Patrick Brady; Chair-Elect: 
Manuela Campanelli; Vice-Chair: Daniel Holz. 
Secretary-Treasurer: James Isenberg; Past Chair:  Steve Detweiler;
Members-at-large:
Scott Hughes, Bernard Whiting,
Laura Cadonati, Luis Lehner,
Michael Landry, Nicolas Yunes.
\parskip=10pt

\vfill
\eject

\vfill\eject

\section*{\centerline
{New developments in space-based gravitational wave astronomy}}
\addtocontents{toc}{\protect\medskip}
\addtocontents{toc}{\bf GGR News:}
\addcontentsline{toc}{subsubsection}{
\it New version of LISA, by Karsten Danzmann}
\parskip=3pt
\begin{center}
Karsten Danzmann, Max Planck Institute for Gravitational Physics
\htmladdnormallink{Karsten.Danzmann-at-aei.mpg.de}
{mailto:Karsten.Danzmann@aei.mpg.de}
\end{center}

For almost two decades now, ESA and NASA have studied the LISA mission
for the observation of low-frequency gravitational waves as an equally
shared partnership of the two agencies.

ESA has recently changed the guidelines for large (``L-class'') missions
in the Cosmic Vision framework to require European-only funding, because
NASA was financially unable to proceed on the timescale of the launch of
the first L-class mission (`L1') in ESA’s Cosmic Vision Programme:

\htmladdnormallink
{\protect {\tt{http://sci.esa.int/science-e/www/area/index.cfm?fareaid=100}}}
{http://sci.esa.int/science-e/www/area/index.cfm?fareaid=100}\\

A search for a European-led variant of LISA that could be launched by
2022 was begun.

After studying several configurations, a new baseline for transfer,
orbit and layout has been identified that will be refined in the coming
month with the help of European industry. The new baseline employs less
costly orbits, and simplifies the design of LISA by reducing the
distance between the satellites and employing four rather than six laser
links. This considerably reduces the mass and cost, while retaining much
of the original science, in part because of new approaches to data analysis.

The European Science Team and a Science Task Force, composed of members
of the gravitational wave and astrophysics communities in both Europe
and the US, have assessed the scientific validity of the new LISA
baseline for the fields of physics, astrophysics and cosmology and have
shown that the new configuration should detect thousands of galactic
binaries, tens of (super)massive black hole mergers out to a redshift of
z=10 and tens of extreme mass ratio inspirals out to a redshift of 1.5
during its two year mission. The investigation of fundamental physics
and cosmology tests will continue over the next few months, until we
have a finalized mission proposal by the fall of 2011. The preliminary
results of this investigation are looking promising.

This announcement is not an official statement of ESA or NASA.

\vfill\eject

\section*{\centerline
{we hear that \dots}}
\addtocontents{toc}{\protect\medskip}
\addcontentsline{toc}{subsubsection}{
\it we hear that \dots , by David Garfinkle}
\parskip=3pt
\begin{center}
David Garfinkle, Oakland University
\htmladdnormallink{garfinkl-at-oakland.edu}
{mailto:garfinkl@oakland.edu}
\end{center}

Daniel Holz was elected Vice Chair of GGR; 
Jim Isenberg was elected Secretary/Treasurer of GGR; and
Michael Landry and Nicolas Yunes were elected Members at large of the Executive Committee of GGR.

Hearty Congratulations!

\vfill\eject

\section*{\centerline
{Finally, Results from Gravity Probe B}}
\addtocontents{toc}{\protect\medskip}
\addtocontents{toc}{\bf Research briefs:}
\addcontentsline{toc}{subsubsection}{
\it Finally, Results from Gravity Probe B, by Clifford M. Will}
\parskip=3pt
\begin{center}
Clifford M. Will, Washington University, St. Louis\footnote{This is an extended version of a ``Viewpoint'' article, published by the American Physical Society in {\em Physics} {\bf 4}, 43 (2011), available at  http://physics.aps.\-org/articles/v4/43.  Published by permission of the APS.}
\htmladdnormallink{cmw-at-wuphys.wustl.edu}
{mailto:cmw@wuphys.wustl.edu}
\end{center}

The great blues singer Etta James' signature song begins, ``At laaasst, my love has come along $\dots$'' This may have been the feeling on May 4, 2011 when NASA announced the long-awaited results of Gravity Probe B \cite{pressconf}, which appeared in Physical Review Letters \cite{prl}. Over 47 years and 750 million dollars in the making, Gravity Probe B was an orbiting physics experiment, designed to test two fundamental predictions of EinsteinÕs general relativity.

%
%
%

According to Einstein's theory, space and time are not the immutable, rigid structures of Newton's universe, but are united as spacetime, and together they are malleable, almost rubbery.  A massive body warps spacetime, the way a bowling ball warps the surface of a trampoline. A rotating body drags spacetime a tiny bit around with it, the way a mixer blade drags a thick batter around.

The spinning Earth does both of these things, and this is what the four gyroscopes aboard the Earth-orbiting satellite Gravity Probe B measured.  The satellite follows a polar orbit with an altitude of 640 kilometers above the Earth's surface. The warping of spacetime exerts a torque on the gyroscope so that its axis slowly precesses -- by about 6.6 arcseconds (or 1.8 thousandths of a degree) per year -- in the plane of the satellite's orbit. (To picture this precession, or ``geodetic effect,'' imagine a stick moving parallel to its length on a closed path along the curved surface of the Earth, returning to its origin pointing in a slightly different direction than when it started.)  The rotation of the Earth also exerts a ``frame-dragging'' effect on the gyro. In this case, the precession is perpendicular to the orbital plane and advances by 40 milliarcseconds per year. Josef Lense and Hans Thirring first pointed out the existence of the frame-dragging phenomenon in 1918, but it was not until the 1960s that George Pugh in the Defense Department and Leonard Schiff at Stanford independently pursued the idea of measuring it with gyroscopes.

There were four gyroscopes aboard Gravity Probe B (GP-B in NASA parlance). Each gyroscope is a fused silica rotor, about the size of a ping-pong ball, machined to be spherical and homogeneous to tolerances better than a part per million, and coated with a thin film of niobium. The gyroscope assembly, which sat in a dewar of 2440 liters of superfluid helium, was held at 1.8 degrees Kelvin. At this temperature, niobium is a superconductor, and the supercurrents in the niobium of each spinning rotor produce a ``London magnetic moment'' parallel to its spin axis.     Extremely sensitive magnetometers (superconducting quantum interference detectors, or ``SQUIDs'') attached to the gyroscope housing are capable of detecting even minute changes in the orientation of the gyros' magnetic moments and hence the precession in their rotation predicted by general relativity.

At the start of the mission, the four gyros were aligned to spin along the symmetry axis of the spacecraft. This was also the optical axis of a telescope directly mounted on the end of the structure housing the rotors. Spacecraft thrusters oriented the telescope to point precisely toward the star IM Pegasi (HR 8703) in our galaxy (except when the Earth intervened, once per orbit). In order to average out numerous unwanted torques on the gyros, the spacecraft rotated about its axis once every 78 seconds.

GP-B started in late 1963 when NASA funded the initial R\&D work that identified the new technologies needed to make such a difficult measurement possible. Francis Everitt became Principal Investigator of GP-B in 1981, and the project moved to the mission design phase in 1984.  Following a major review of the program by a National Academy of Sciences committee in 1994, GP-B was approved for flight development, and began to collaborate with Lockheed-Martin and Marshall Space Flight Center. The satellite launched on April 20, 2004 for a planned 16-month mission, but another five years of data analysis were needed to tease out the effects of relativity from a background of other disturbances of the gyros.   

Almost every aspect of the spacecraft, its subsystems, and the science instrumentation performed extremely well, some far better than expected. Still, the success of such a complex and delicate experiment boils down to figuring out the sources of error. In particular, having an accurate calibration of the electronic readout from the SQUID magnetometers with respect to the tilt of the gyros was essential. The plan for calibrating the SQUIDs was to exploit the aberration of starlight, which causes a precisely calculable misalignment between the rotors and the telescope as the latter shifts its pointing toward the guide star by up to 20 arcseconds to compensate for the orbital motion of the spacecraft and the Earth. However, three important, but unexpected, phenomena were discovered during the experiment that affected the accuracy of the results.

%
%
%

First, because each rotor is not exactly spherical, its principal axis rotates around its spin axis with a period of several hours, with a fixed angle between the two axes.  This is the familiar ``polhode'' period of a spinning top.  In fact this polhoding was essential in the calibration process because it led to modulations of the SQUID output via the residual trapped magnetic flux on each rotor (about 1 percent of the London moment).   But the polhode period and angle of each rotor actually decreased monotonically with time, implying the presence of some damping mechanism, and this significantly complicated the calibration analysis. In addition, each rotor was found to make occasional, seemingly random ``jumps'' in its orientation -- some as large as 100 milliarcseconds. Some rotors displayed more frequent jumps than others. Without being able to continuously monitor the rotors' orientation, Everitt and his team couldn't fully exploit the calibrating effect of the stellar aberration in their analysis.  Finally, during a planned 40-day, end-of-mission calibration phase, the team discovered that when the spacecraft was deliberately pointed away from the guide star by a large angle, the misalignment induced much larger torques on the rotors than expected. From this, they inferred that even the very small misalignments that occurred during the science phase of the mission induced torques that were probably several hundred times larger than the designers had estimated.

\begin{table}[t]
\begin{center}
\begin{tabular}{ccc}
\hline
&Measured&Predicted\\
\hline
\\
Geodetic Precession (mas)&$6602 \pm 18$&$6606$\\
\\
Frame-dragging (mas)&$37.2 \pm  7.2$&$39.2$\\
\\
\hline
\end{tabular}
\caption{ Final results of Gravity Probe B}
 \label{tbl-1}
\end{center}
\end{table}

What ensued during the data analysis phase was worthy of a detective novel. The critical clue came from the calibration tests. Here, they took advantage of the residual trapped magnetic flux on the gyroscope. (The designers used superconducting lead shielding to suppress stray fields before they cooled the niobium coated gyroscopes, but no shielding is ever perfect.) This flux adds a periodic modulation to the SQUID output, which the team used to figure out the phase and polhode angle of each rotor throughout the mission. This helped them to figure out that interactions between random patches of electrostatic potential fixed to the surface of each rotor, and similar patches on the inner surface of its spherical housing, were causing the extraneous torques.  In principle, the rolling spacecraft should have suppressed these effects, but they were larger than expected. 

Fortunately, the patches are fixed on the various surfaces, and so it was possible to build a parametrized model of the patches on both surfaces using multipole expansions, and to calculate the torques induced by those interactions when the spin and spacecraft axes are misaligned, as a function of the parameters.   One prediction of the model is that the induced torque should be perpendicular to the plane formed by the two axes, and this was clearly seen in the data.  Another prediction is that, when the slowing decreasing polhode period crosses an integer multiple of the spacecraft roll period, the torques fail to average over the roll period, whereupon the spin axis precesses about its initial direction in an opening Cornu spiral, then migrates to a new direction along a closing Cornu spiral.  This is known as a loxodromic path, familiar to navigators as a path of fixed bearing on the Earth's surface.  Detailed observation of the orientation of the rotors during such ``resonant jumps'' showed just such loxodromic behavior.  In the end, every jump of every rotor could be identified by its ``mode number'', the integer relating its polhode period to the spacecraft roll period.   

The original goal of GP-B was to measure the frame-dragging precession with an accuracy of 1\%, but the problems discovered over the course of the mission dashed the initial optimism that this was possible. Although Everitt and his team were able to model the effects of the patches, they had to pay the price of the increase in error that comes from using a model with so many parameters. The experiment uncertainty quoted in the final result -- roughly 20\% for frame dragging -- is almost totally dominated by those errors. Nevertheless, after the model was applied to each rotor, all four gyros showed consistent relativistic precessions. Gyro 2 was particularly ``unlucky'' -- it had the largest uncertainties because it suffered the most resonant jumps.  Numerous cross-checks were carried out, including estimating the relativity effect during different segments of the 12-month science phase (various events, including computer reboots and a massive solar storm in January 2005, caused brief interruptions in data taking), increasing and decreasing the number of parameters in the torque model, and so on. 

When GP-B was first conceived in the early 1960s, tests of general relativity were few and far between, and most were of limited precision. But during the ensuing decades, researchers made enormous progress in experimental gravity, performing tests of the theory by studying the solar system and binary pulsars \cite{wer}.  Already by the middle 1970s, some argued that the so-called parametrized post-Newtonian (PPN) parameters that characterize metric theories of gravity, like general relativity, were already known to better accuracy than GP-B could ever achieve \cite{livrev}. Given its projected high cost, critics argued for the cancellation of the GP-B mission. The counter-argument was that all such assertions involved theoretical assumptions about the class of theories encompassed by the PPN approach, and that all existing bounds on the post-Newtonian parameters involved phenomena entirely different from the precession of a gyroscope.  All these issues were debated, for example, in the 1994 NAS/NRC review of GP-B that recommended its continuation.

The most serious competition for the results from GP-B comes from the LAGEOS experiment, in which laser ranging accurately tracked the paths of two laser geodynamics satellites orbiting the Earth. Relativistic frame dragging was expected to induce a small precession (around 30 milliarcseconds per year) of the orbital plane of each satellite in the direction of the Earth's rotation. However, the competing Newtonian effect of the Earth's nonspherical shape had to be subtracted to very high precision using a model of the Earth's gravity field. The first published result from LAGEOS in 1998 \cite{lageos1,lageos2} quoted an error for the frame-dragging measurement of 20 to 30\%, though this result was likely too optimistic given the quality of the gravity models available at the time. Later, the GRACE geodesy mission offered dramatically improved Earth gravity models, and the analysis of the LAGEOS satellites finally yielded tests at a quoted level of approximately 10\% \cite{lageos3}.

Frame dragging has implications beyond the solar system. The incredible outpouring of energy from quasars along narrow jets of matter that stream at nearly the speed of light is most likely driven by the same frame-dragging phenomenon measured by GP-B and LAGEOS.  In the case of quasars, the central body is a rapidly rotating black hole.  In another example, the final inward spiral and merger of two spinning black holes involve truly wild gyrations of each body's spin axes and of the orbit, again driven by the same frame-dragging effect, and these motions are encoded in gravitational-wave signals.  Laser interferometric observatories on the ground, and in the future, a similar observatory in space, may detect these gravity waves. So there is a strong link between the physics Gravity Probe B was designed to uncover and that describing some of the most energetic and cataclysmic events in the universe.

Even though it is popular lore that Einstein was right (I even wrote a book on the subject), no such book is ever completely closed in science. As we have seen with the 1998 discovery that the universe is accelerating, measuring an effect contrary to established dogma can open the door to a whole new world of understanding, as well as of mystery.  The precession of a gyroscope in the gravitation field of a rotating body had never been measured before GP-B.  While the results support Einstein, this didn't have to be the case.  Physicists will never cease testing their basic theories, out of curiosity that new physics could exist beyond the ``accepted'' picture.

\medskip
\noindent
{\em Disclosure: CMW chaired NASA's external Science Advisory Committee for Gravity Probe-B from 1998 to 2011.}

\vfill\eject

\section*{\centerline
{Discovery and EOS implications}\\ 
\centerline {of the highest-mass ($2\,\Msun$)
neutron star}}
\addtocontents{toc}{\protect\medskip}
\addcontentsline{toc}{subsubsection}{
\it Discovery of the highest-mass 
neutron star, by Wynn C.G. Ho}
\parskip=3pt
\begin{center}
Wynn C.G. Ho, University of Southampton
\htmladdnormallink{wynnho-at-soton.ac.uk}
{mailto:wynnho@soton.ac.uk}
\end{center}

In October 2010, radio astronomers announced the mass measurement of the
neutron star known as PSR~J1614$-$2230 \cite{demorestetal10}.
The neutron star mass $M=1.97\pm0.04\,\Msun$ is the most-accurate, high-mass
neutron star known to date.
PSR~J1614$-$2230 is a radio pulsar with a 3.15~ms spin period and is in a
8.7~d binary orbit with a $M_{\mathrm{c}}=0.5\,\Msun$ companion star.
What contributed to making the mass measurement possible is that the binary
system is viewed nearly edge-on, with an orbital inclination angle
$i=89^\circ.17\pm 0^\circ.02$.
The edge-on view allows a measurement of Shapiro delay and,
combined with the large companion mass and excellent timing precision of a new
instrument on the National Radio Astronomy Observatory Green Bank Telescope,
yields a very accurate measurement of the mass of a neutron star.

Shapiro delay is a delay in the measured arrival times of light
pulses from the neutron star due to this light passing through the
gravitational potential of the companion star \cite{shapiro64}.
For nearly circular orbits, Shapiro delay is
$\Delta_{\mathrm{S}}\approx -2r\ln(1-s\sin\Phi)$,
where $\Phi$ is the orbital phase and $r$ and $s$ are the Shapiro delay
range and shape parameters, respectively, and are given by
$r=GM_{\mathrm{c}}/c^3=4.9\mbox{ $\mu$s }(M_{\mathrm{c}}/\Msun)$
and $s=\sin i$ \cite{damourtaylor92,lorimerkramer05}.
When the effect of Shapiro delay is strong, timing residuals from the pulsar
signal show a characteristic peak at orbital phase $\Phi=90^\circ$, which
corresponds to superior conjunction or when the pulsar is behind companion
star.  Other measurements, e.g., gravitational redshift, orbital decay due
to gravitational radiation, and orbital precession, allow degeneracies
in the Shapiro delay equation to be broken and the mass to be determined
uniquely and accurately.
See \cite{krameretal06} for a demonstration of this technique as applied to the
double pulsar system PSR~J0737$-$3039A/B, which yields the very accurate
(uncertainty in last digit given in parentheses) neutron star masses
$1.3381(7)\,\Msun$ and $1.2489(7)\,\Msun$.
Note that PSR~J0737$-$3039A/B has an inclination angle of
$88^\circ.69^{{+0^\circ.50}}_{{-0^\circ.76}}$, so is viewed (marginally)
less edge-on than PSR~J1614$-$2230.
Note also that the 3.15~ms spin period of PSR~J1614$-$2230 is too slow
to pose interesting constraints.
 
One of the reasons for studying neutron stars is to use them as probes of
fundamental physics in regimes that cannot be accessed in laboratories on
Earth.  An important example is the nuclear equation of state (EOS).
So what is the nuclear EOS, and how does measuring the mass of a
neutron star constrain the EOS?
I summarize arguments made in \cite{lattimerprakash10}; see for details.
The nuclear EOS determines the behavior of matter near nuclear densities
(at $n_{\mathrm{nuc}}\approx 0.16\mbox{ fm$^{-3}$}$ or
$\rho_{\mathrm{nuc}}\approx 2.8\times 10^{14}\mbox{ g cm$^{-3}$}$)
and provides a relationship between pressure and density, i.e., $P(\rho)$.
While the EOS is fairly well-known at $\rho\ll\rho_{\mathrm{nuc}}$,
large uncertainties exist at $\rho\gtrsim\rho_{\mathrm{nuc}}$, and
there are many detailed theoretical calculations of what it might be
(see, e.g., \cite{lattimerprakash01,lattimerprakash07}, and references therein).
The EOS also determines the abundances of particles that comprise a neutron
star; in the core near nuclear densities, the star is composed of neutrons,
protons, electrons, and a small amount of muons.
If the deep central regions of the neutron star can sustain high enough
densities, a lower energy ground state may be achieved by the formation of
particles beyond the typical constituents, exotic particles such as kaons,
hyperons, or deconfined (up, down, and strange) quarks
(see, e.g., \cite{lattimerprakash04}, for review).

How does the EOS affect neutron star mass (and radius)?
In order to build a model of the structure of a cold, spherical
(relativistic) star, one solves the Tolman-Oppenheimer-Volkoff (TOV) equations
(\cite{tolman39,oppenheimervolkoff39}; see also \cite{shapiroteukolsky83}):
these equations describe hydrostatic equilibrium, mass conservation, and
gravitational acceleration and
are first-order differential equations for pressure $P$, enclosed
mass $m$, and gravitational potential $\Phi$ as a function of radial
distance $r$.
An EOS $P(\rho)$ is needed to close this set of equations.
The inner boundary condition, central density $\densc$ [$\equiv\rho(r=0)$,
which also prescribes the central pressure $P_{\mathrm{c}}$ via the EOS],
is chosen.
The TOV equations are then integrated outward until $P=0$, and
this defines the total mass $M$ and radius $R$ of the star, i.e.,
$R\equiv r(P=0)$ and $M\equiv m(R)$.
For a given theoretical $P(\rho)$, a $M$-$R$ sequence can be constructed.
The maximum mass $\Mmax$ neutron star in the sequence is built using the
maximum allowed central density $\densmax$ for that EOS.
A star that is more massive than $\Mmax$ would form a black hole.
Note that General Relativity and causality
(sound speed everywhere in the star is less than the speed of light;
\cite{rhoadesruffini74,lindblom84,lattimerprakash04})
provide constraints on $M/R$,
i.e., $2GM<c^2R$ and $3GM<c^2R$, respectively.
Normal matter EOSs yield stable configurations whose radius decreases with
increasing mass, while quark EOSs produce
``quark'' or ``strange'' stars \cite{witten84,farhijaffe84}
whose radius increases with increasing mass.
EOSs that are ``stiff/soft'' are ones which produce higher/lower $\Mmax$
for a given $R$, and EOSs that involve exotic particles are generally soft
EOSs with lower $\Mmax$.

This last point highlights why it is important to measure the maximum mass
of neutron stars.
If the measured $\Mmax$ is higher than the maximum mass allowed by a
given theoretical EOS, then that EOS is ruled out.
Taking $2\,\Msun$ to be the lower limit of the maximum mass, i.e.,
$\Mmax\ge 2\,\Msun$, does this rule out any interesting EOSs?
Unfortunately, the answer is no.
Neutron star models containing, e.g., kaons, have been constructed with
$\Mmax\sim 2\,\Msun$ (see \cite{lattimerprakash10}, and references therein).
For quark stars, \cite{witten84} showed that $\Mmax=2.5\,\Msun$ is the
absolute limit; more realistic quark star models have lower values of
$\Mmax$, but these can be above $2\,\Msun$.
A reliable elimination of soft EOSs, such as those that produce strange
quark stars, would be possible with a measurement of a $2.4\,\Msun$
neutron star
(intriguingly the black widow pulsar, PSR~B1957+20, has $M=2.4\,\Msun$ though
with rather large uncertainty \cite{reynoldsetal07,vankerkwijketal11}).
Nevertheless, the measured (highest) neutron star mass does constrain
\densmax,
which is given by 
$\densmax\approx 1.4\times 10^{16}\mbox{ g cm$^{-3}$}(\Msun/\Mmax)^2$,
for all EOSs \cite{lattimerprakash10}.
With $M=2\,\Msun$ for PSR~J1614$-$2230, the central density for any
neutron star must be less than $3.5\times 10^{15}\mbox{ g cm$^{-3}$}$.

\vfill\eject

\section*{\centerline
{Numerical relativity beyond astrophysics}}
\addtocontents{toc}{\protect\medskip}
\addtocontents{toc}{\bf Conference reports:}
\addcontentsline{toc}{subsubsection}{
\it Numerical relativity beyond astrophysics, 
by Carsten Gundlach}
\parskip=3pt
\begin{center}
Carsten Gundlach, University of Southampton 
\htmladdnormallink{c.j.gundlach-at-soton.ac.uk}
{mailto:c.j.gundlach@soton.ac.uk}
\end{center}

A workshop on ``Numerical relativity beyond astrophysics" was held at the
International Centre for Mathematical Sciences (Edinburgh), 11-15
July 2011.  (Here ``beyond'' is just funding body-speak for ``not in''.) The
meeting was intended to refocus a bit on those applications of
numerical relativity that are not motivated by astrophysical
applications or the search for gravitational waves: critical collapse,
higher dimensional gravity, AdS-CFT, etc. It was organised jointly by
Carsten Gundlach, David Garfinkle and Luis Lehner. We tried to invite a mix of
``numerical'', ``mathematical'' and ``strings'' people, and focus each
day on a particular topic, supplemented by a daily discussion session
in the long lunchbreak.

Monday saw a good lineup of talks on {\bf critical collapse}, with
reviews of older work by {\it Gundlach}, {\it Oliver Rinne} and {\it
  Steve Liebling}. {\it Evgeny Sorkin} talked about his simulations of
vacuum critical collapse in axisymmetry. He has not yet been able to
repeat the results of Abrahams and Evans 1992, but claims to see a
different kind of discrete self-similarity at the black hole
threshold, with the maximum of the curvature on a ring. If this holds
up, it would indicate discrete self-similarity in cylindrical symmetry
(with the ring appearing straight at small scales). 

The discussion session was a round-robin of introductions.

Tuesday was dedicated mainly to {\bf black holes}. {\it Toby Wiseman}
talked about methods for finding the zillions of black hole solutions
in higher dimensions. If they are static, Wick rotation gives elliptic
equations, which can be solved for example using Ricci flow. If they
are only stationary, then more brute force methods are required. {\it
  Jorge Santos} talked about black holes with a single Killing vector
field (in the first instance, in AdS5, with scalar hair). The KV has
to be normal to the horizon to avoid rigidity theorems. {\it Lehner}
reviewed recent work that settles the fate of the famous
Gregory-Laflamme instability of a black string. Any almost-pinched off
part of the string experiences a new GL instability, and so on,
resulting in a naked singularity in finite time. {\it Fethi
  Ramazanoglu} talked about numerical simulations of black hole
evaporation in 1+1 dimensions in the CGHS toy model of quantum
gravity. 

The discussion session was on critical collapse. It is very much still
to be understood in axisymmetry!

Wednesday continued the black holes theme, with a review talk by
{\it Mihalis Dafermos} on the effort to prove nonlinear
stability. {\it Pat Brady} gave a review on black hole interiors: mass
inflation for charged spherical black holes is now well understood,
but Kerr is very much an open problem. The physical interpretation of
the weak null singularity that precedes the usual spacelike one is
also still unclear. {\it Masaru Shibata} showed impressive numerical
evolutions of black holes in $D=5,6,7$ dimensions (assuming $SO(D-3)$
symmetry). The Myers-Perry black hole is unstable above a critical
spin, as expected. {\it Miguel Zilhao} and {\it Helvi Witek} gave a
joint talk on recent simulations of BH collisions in higher dimension,
again in $SO(D-3)$ symmetry. {\it Amos Ori} followed up Brady's talk
with some speculation on what one expects to see in Kerr interiors and
suggestions for numerical simulations. -- The discussion session was on
AdS-CFT. It began with an informal presentation by Wiseman, which led
to very lively discussion.

Thursday was dedicated to {\bf AdS-CFT}, but this strand had in effect
begun on Tuesday afternoon with a lovely review talk for
non-specialists by {\it Andrei Starinets}. {\it Piotr Bizon} presented
strong numerical and analytical evidence for a nonlinear instability
of AdS with a minimally coupled scalar field (with Dirichlet boundary
conditions). He explains it as resonance phenomenon in third-order
perturbation theory made possible by the fact that the (linear)
spherical scalar field modes have frequencies that differ by
integers. This mechanism works in 3+1 and higher dimensions. It is
slow, but much faster than the instability one might expect from
Poincar\'e recurrence for a system in a box.  {\it Jorma Louko}
reviewed one key application of AdS-CFT, to the thermalisation of
expanding plasmas (as a toy model for high-energy collisions of nuclei
at RHIC). {\it Frans Pretorius} presented a general numerical method
for evolving 5D asymptotically AdS spacetimes, using scalar matter
fields and the generalised harmonic formulation. {\it Nick Evans}
talked about probe branes as models of quarks at finite temperature,
and {\it Harvey Reall} about horizon instabilities and local Penrose
inequalities.

The extremely lively discussion session was a continuation of the
previous one on AdS-CFT, this time kicked off by an informal
presentation by Bob Wald querying the notion of causality in
AdS-CFT. The way the string theorists think about the gravity side of
AdS-CFT is certainly not the way relativists naturally think about an
initial-boundary value problem for asymptotically AdS spacetimes!
From gravity in the bulk one can read off a (conserved) stress-energy
tensor on the boundary, which is supposed to be a quantum expectation
value. But how does the quantum state in the boundary theory relate to
classical initial data in the bulk? 

A second discussion took place in the afternoon, on black hole
interiors, kicked off by an informal presentation by Dafermos.

On Friday {\it Wald} discussed the bobbing and kicks seen in binary
black hole mergers in terms of special relativistic toy models. These
can be made rigorous in GR by introducing suitable center of mass
world line and frame. {\it Lee Lindblom} presented a general method for solving
PDEs on manifolds with arbitrary spatial topology. {\it Olivier
  Sarbach} reviewed the Cauchy problem for the Einstein equations on a
finite domain, a topic pioneered by Friedrich for asymptotically AdS
spacetimes before AdS-CFT became fashionable, and now relevant for
numerical relativity. {\it Garfinkle} gave the last talk, on numerical
simulations of the collapse of k-essence (a scalar field with
nonstandard kinetic energy in the Lagrangian). 

A hardy bunch met for tea and a last discussion session.  The meeting
had been rounded out by communal lunches at ICMS, two conference
dinners, and even communal Scottish breakfast for those who felt like
talking shop so early.

\vfill\eject

\section*{\centerline
{ICTP workshop on ``Cold Materials, Hot Nuclei and Black Holes"}}
\addtocontents{toc}{\protect\medskip}
\addcontentsline{toc}{subsubsection}{
\it Cold Materials, Hot Nuclei and Black Holes, 
by Vijay Balasubramanian}
\parskip=3pt
\begin{center}
Vijay Balasubramanian, University of Pennsylvania
\htmladdnormallink{vijay-at-physics.upenn.edu}
{mailto:vijay@physics.upenn.edu}
\end{center}

	Between the 15th and 26th of August this year, the ICTP, Trieste hosted a workshop focused on applications of the AdS/CFT correspondence to strongly coupled QCD and condensed matter systems. The workshop was organized by Vijay Balasubramanian, Jan de Boer,
Veronika E.\ Hubeny, and Mukund Rangamani.  There were two talks each morning and in the afternoon participants broke up into small groups discussing diverse topics.  The discussions lasted through dinner, and then started again, often continuing past midnight on a patio facing the Adriatic Sea.  It was just like a physics summer camp, complete with participants jumping into the nearby water to cool off.

	The talks on applications to QCD focused primarily on the problem of thermalization and entropy generation.   At RHIC, and more recently at the LHC, it has been shown that heavy ion collisions produce a complicated state of matter which thermalizes rapidly and produces a lot of entropy until it becomes a quark-gluon plasma which evolves hydrodynamically till the point where it hadronizes.  Nuclear theorist Berndt Muller gave a talk discussing the extensive data that is becoming available, and presented the string theorists with various challenges (e.g. explaining the fluctuations in the energy emitted at different angular separations).  Derek Teaney, also a nuclear theorist, explained how real-time methods from quantum field theory could be adapted to AdS/CFT to calculate properties of non-equilibrium configurations. Diana Vaman discussed new ways to perform real-time AdS/CFT computations and their applications to jet quenching, and Piljin Yi gave an overview of holographic approaches to studying baryons.  Finally, Ben Craps described various holographic probes of thermalization and how they react in a model where energy is injected suddenly and then proceeds to equilibrate.

	Several talks focused on the states of matter that are well approximated by fluid dynamics and their holographic description.   Amos Yarom discussed holographic superfluids, Ramalingam Loganayagam discussed a holographic approach to the Wilsonian renormalization group that has a bearing on the fluid/gravity correspondence,  and Cindy Keeler discussed the Harvard group's approach to getting a fluid dynamical description of gravity.  A substantial number of speakers addressed the subtle issues in getting Fermi liquids and non-Fermi liquids in the AdS/CFT correspondence.  Sandip Trivedi discussed how a variety of dilaton coupled gravities would give rise to different kinds of Fermi liquid behaviors.  Shamit Kachru discussed instabilities of the AdS$_2$ space that typically appears in the near-horizon of the extremal black holes used to model cold materials, and explained how these instabilities will modify the deep infrared behavior of the dual field theory.  Jerome Gauntlett discussed top-down constructions yielding interesting charged liquids and both Larus Thorlacius and Koenraad Schalm discussed how the back-reaction of fermions affects the AdS/CFT constructions of Fermi liquids.

	Several talks addressed fundamental issues in the AdS/CFT correspondence that have a direct bearing on applications to condensed matter systems and QCD.  Andrei Parnachev discussed conformal phase transitions at strong and weak coupling, and Simon Ross discussed the holographic description of asymptotically Lifshitz universes where space and time scale differently with the radius in spacetime.  Micha Berkooz discussed the construction of D-brane configurations which could realize the sorts of intersecting defects that occur in many interesting condensed matter problems.  Rob Myers gave a beautiful discussion of the Ryu-Takayanagi formula for holographic entanglement entropy and discussed approaches to deriving this formula from first principles, at least in some special cases.
	
	Two provocative talks, one in the first week, and one in the second, addressed foundational issues.   The first, by the condensed matter theorist Sung-Sik Lee, proposed a novel characterization of gauge theories with gravity duals.  The idea involved studying the dynamics of Wilson loops at certain critical points.  Conference participants found this talk and the perspective it advanced very stimulating.   The second talk, by Joan Simon, considered the problem of fast scrambling (the idea that black holes scramble information as fast as possible without violating any laws of physics).  Joan discussed possible dynamical models that would give rise to this fundamental property which is implicated in all the studies that use AdS/CFT to study thermal behavior and thermalization.
		
	The talks were uniformly of a very high quality.  This was a very stimulating and successful meeting.  In between discussions some people ate horse and wild boar.  No one tried the spiced mountain goat.   And no one was attacked viciously by sharks, bears, or even other participants.

\vfill\eject

\section*{\centerline
{Benasque Workshop on Gravity}}
\addtocontents{toc}{\protect\medskip}
\addcontentsline{toc}{subsubsection}{
\it Benasque Workshop on Gravity, 
by \'Oscar J. C. Dias}
\parskip=3pt
\begin{center}
\'Oscar J. C. Dias, DAMTP, University of Cambridge
\htmladdnormallink{O.Dias-at-damtp.cam.ac.uk}
{mailto:O.Dias@damtp.cam.ac.uk}
\end{center}

Benasque is a charming little village in the heart of the Spanish Pyrenees, north of Barcelona.
During 15 years, several scientific meetings took place in this
village, mainly as the result of the effort undertaken by the
Spanish physicist Pedro Pascual (1934-2006). As a recognition of his
initiative, this  Center for Science is now called ``The Centro de
Ciencias de Benasque Pedro Pascual". The Summer of 2009 signaled the
opening of its new building, a modern facility designed as a place
for work and interaction: plenty of desk space, omnipresent
blackboards, common areas with coffee machines, and multifunctional
conference rooms. This date and venue also marked the realization of
the workshop  ``Gravity: New perspectives from strings and higher
dimensions". Given the success of this experience (and as demanded
by several of the participants!), the second edition of this
workshop took place during July 17-29, 2011 (organized by Roberto
Emparan, Veronika Hubeny and Mukund Rangamani) .

Over the last decade we have had remarkable progress in extending
the field of application of General Relativity and in understanding
various aspects of higher-dimensional gravity. Not only have we
started to see many intriguing new solutions, but we are also
witnessing the use of classical gravity as a tool being applied to
plasma physics and condensed matter systems, and big advances in
numerical relativity. More than fifty participants, gathered to
discuss these developments, the current progress, and the future
directions. The scheduled program for the workshop was light, with
two hours of talks per day, plus a number of impromptu discussion
sessions during some of the evenings (trust me: a lot of enthusiasm
and action here!). There was time to enjoy the marvelous hiking in
the surrounding mountains, and the local food. Some of the
participants (and family) also extended their trip and visited
Barcelona.


Mukund Rangamani opened the workshop with a discussion on the
formulation of a holographic Wilsonian renormalization group flow
for strongly coupled systems with a gravity dual. This provides an
efficient extraction of low energy behavior of the system. The idea
is to start  with field theories defined on a cut-off surface in a
bulk spacetime. Then, integrating out high energy modes in the field
theory corresponds to integrating out a part of the bulk geometry.
This formulation can be used to derive a semi-holographic
description of low energy physics and provides an AdS/CFT
interpretation of the membrane paradigm. (Work with Faulkner, Liu;
Bratton, Camps, Loganayagam).


Roberto Emparan described the current state-of-the-art  in what
concerns the phase diagram of stationary, vacuum, asymptotically
flat black holes in higher dimensional spacetimes. A full
classification of these solutions is challenging, and he told us how
the blackfold approach is a successful technique to identify
important patterns in the phase diagram. This blackfold approach is
an effective worldvolume theory for the dynamics of black branes. It
allows for the perturbative construction of new black hole
solutions. In addition, it is useful for the analysis of dynamical,
non-stationary situations, like instabilities of black branes.
(Work with Harmark, Niarchos, Obers;  Caldarelli, Camps, Haddad,
Rodriguez).


Jorge Santos presented the first examples  of AdS black holes  with
scalar hair that are invariant under a single Killing field, which
is the null generator of the horizon (so, they are neither stationary
nor axisymmetric). They are related to rotating boson stars and
to the endpoint of a superradiant instability. Moving to a related
topic, Jorge reminded us that  AdS spacetime with a scalar field is
non-linearly unstable to transfering energy to smaller and smaller
scales and eventually forming a small black hole. He then told us
that a gravitational turbulent instability is also present if we try
to construct non-linear geons. The implications of this turbulent
instability for the dual field theory, and the existence of single
Killing field black hole counterparts of the geon were discussed.
(Work with Dias, Horowitz).


Toby Wiseman reviewed the numerical framework for finding  static
and stationary vacuum black hole solutions in higher dimensions. He
first explained the advantages of solving the Harmonic Einstein
equation instead of the usual vacuum Einstein equation: the former
is an explicit elliptic system. We can solve such a system using two
relaxation methods, namely the Ricci flow (local relaxation) and
Newton's methods. Toby then described two explicit solutions
constructed using these elliptic numerical methods. One is the
$AdS_5$ classical gravity dual of the CFT$_4$ on a Schwarzschild
boundary background in the Unruh vacuum. The other is a large Randall
Sundrum II  static black hole.
 (Work with  Headrick, Kitchen; Figueras, Lucietti).


Robert Myers gave a derivation of holographic entanglement entropy
for spherical entangling surfaces. It relies on conformally mapping
the  boundary CFT to a hyperbolic geometry and observing that the
vacuum state is mapped to a thermal state in the latter geometry. He
also  discussed holographic entanglement entropy with higher
curvature gravity in the bulk. Here, Wald's formula for horizon
entropy does not yield the correct entanglement entropy. However,
for Lovelock gravity, Rob told us that there is an alternate
prescription which involves only the intrinsic curvature of the bulk
surface.
 (Work with Hung, Smolkin; Casini, Huerta).


Axions are scalars taking values in a circle; their periodicity
prevents perturbative corrections to the potential; hence their mass
is generated by nonperturbative effects. In confining gauge
theories, one often  finds the axion monodromy phenomenon: the
energy is  a multivalued functional of axion's angle, with a tower
of metastable states above the ground state. The spectrum is
periodic in shifts of the polar angle $\theta$ by $2\pi$ but the
different states mix and reshuffe, and the energy grows with
$\theta$.  Albion  Lawrence discussed how axion monodromy can lead
to interesting phenomena in cosmology and astrophysics, and how one
can use it  to build inflation models in field theory and string
theory.  (Work with Kaloper, Sorbo; Dubovsky, Roberts).


Donald Marolf discussed the encouraging  ongoing efforts to find the
$AdS_5$ classical gravity dual of the CFT$_4$ on a Schwarzschild
boundary background in the Hartle-Hawking vacuum. This will be the
AdS/CFT dual of Hawking radiation. More concretely, he described how
the strong coupling behaviour of quantum field theories on a
non-dynamical boundary black hole background can be described, in
the context of the AdS/CFT correspondence, by a competition between
two gravity duals: a black funnel and a black droplet. In this
context, Don also told us about the possible existence of ``flowing
black funnels" that have a horizon rotating with different
velocites as we move from the boundary into the bulk (these
solutions have no Killing horizon so the rigidity theorem does not
apply).  (Work with Hubeny, Rangamani; Raamsdonk; Santos, Way; Fischetti).


Vitor Cardoso highlighted the window of opportunities that the
gravitational  wave detector experiments will provide for
gravitational wave physics/astrophysics, for the strong-curvature
regime of gravity, for black hole spectroscopy, and to test
alternative theories of gravity. He reviewed the numerical evolution
studies of the black hole binary/coalescence system: its three main
phases (inspiral, merger and ring-down) and the technique used to
extract the physical signal. Black hole collisions provide the most
energetic phenomena in the universe,  the best source of
gravitational wave emission and they test the cosmic censorship and
hoop conjectures. Several time evolution codes are now available and
Vitor told us that we can study high energy head-on and grazing
collisions in four and higher dimensions. It is  also possible to
simulate zoom-whirl collisions where the black holes scatter after
orbiting several times around each other, and collisions with kicks
where the final black black is ejected with a large linear velocity.


In the last two years it was found that asymptotically flat, vacuum,
black holes are unstable to the ultraspinning and  bar-mode
instabilities if they have large angular momenta. These results rely
on numerical studies. Harvey Reall described how we can use simpler
analytical methods to prove the existence  of certain types of
horizon instabilities. The idea is to find initial data that
describes a small perturbation of the black hole and violates a
local Penrose inequality. He told us how this approach confirms the
existence of the  Gregory-Laflamme instability on a black string and
the existence of the ultraspinning instability. Moreover, it also
proves that ``fat" black rings are unstable.  (Work with Figueras,
Murata).


Stefano Giusto gave a detailed survey of the generating techniques
that  are being used to find extremal but non-BPS multi-center
solutions, following their successful application to BPS systems. We
need these solutions if we want to understand  black holes within
string theory and the microscopic origin of their Bekenstein-Hawking
entropy. This program is particularly relevant to find if, and in
what form, the so-called fuzzball proposal applies to BPS and
non-BPS black holes.  This proposal claims the existence of smooth
horizonless configurations, with the same charges of the macroscopic
black hole, that would be the microstates of the system  and account
for the statistical description of  black hole thermodynamics.
(Work with Bena, Bobev, Dall'Agata, Ruef, Warner).


Valeri Frolov considered  four dimensional black holes embedded in models with large
extra dimensions, and how events which are causally-disconnected along a lower dimensional
hypersurface may be causally connected in the full spacetime.
The idea is to study  an induced geometry on a
test brane in the background of a black string/brane. At the
intersection surface of the test brane with the bulk black
string/brane the induced metric has an event horizon; so the test
brane contains a black hole dubbed a ``brane hole". If the test
brane moves with respect to the bulk, the emission and absorption of
photons can be used to learn about the black hole interior. Indeed,
from a test brane viewpoint such events are connected by a spacelike
curve in the induced geometry.  (Work with Mukohyama; Gorbonos).
(During this workshop, Valeri also took the opportunity to promote
his most recent book that includes a description of the more recent
developments in black hole physics. The reader will certainly enjoy
it!)


Witten diagrams are the tool for calculating correlation functions
of  strongly coupled conformal field theories with a gravity dual.
In spite of significant progress, these calculations are however not
easy and only a small number of explicit computations have been done.
Traditionally, these are performed either in coordinate or in
momentum spaces. Miguel Paulos proposed changing basis. He told us
what are the benefits of  working with the  embedding formalism and
the Mellin transform. This allows the calculation of certain
tree-level conformal correlation functions in AdS/CFT that were not
computed using the standard Fourier transform.


Take the QCD phase diagram of temperature $T$ versus chemical
potential  $\mu$ for baryon number. There are computational tools
available to study this system in two regions, namely the large $T$
(QGP) regime and the large $\mu$  (CFL) phase. In between, there is
a strongly coupled region that should be described by strongly
interacting deconfined quarks. Unfortunately, the Lattice QCD
simulations are not a useful tool to explore it. Prem Kumar
proposed using a dual holographic gravity description  of the system
to get insights about the fundamental properties of this system. The
idea is to use D-brane systems that have a confinement/deconfinement
phase transition, and a phase diagram that shares important common
features with QCD, to learn about strongly coupled quark dense
matter. This program is being implemented in the D3/D7 brane system
and associated Hedgehog black hole. (Work with Benicasa).

During the workshop, there were enthusiastic discussions and intense interactions.
The overall open and collaborative atmosphere was very much appreciated by
everyone. It was generally felt that such a successful and enjoyable meeting should have a
continuation, and plans to hold a third Benasque Workshop on Gravity in July
2013 are already underway.


\begin{thebibliography}{99}

\bibitem{pressconf}
Press conference available at 
\htmladdnormallink
{\protect {\tt{http://www.youtube.com/\-watch?v=SBiY0Fn1ze4}}}
{http://www.youtube.com/watch?v=SBiY0Fn1ze4}\\


\bibitem{prl}
C. W. F. Everitt et al., Phys. Rev. Lett. 106, 221101 (2011).

\bibitem{wer}
C. M. Will, Was Einstein Right? (Basic Books, Perseus, NY, 1993).

\bibitem{livrev}
C. M. Will, Living Rev. Relativ. 9, 3 (2006); 
\htmladdnormallink
{\protect {\tt{http://www.livingreviews.org/lrr-2006-3}}}
{http://www.livingreviews.org/lrr-2006-3}\\


\bibitem{lageos1}
I. Ciufolini et al., Class. Quantum Gravit. 14, 2701 (1997).

\bibitem{lageos2}
I. Ciufolini et al., Science 279, 2100 (1998).

\bibitem{lageos3}
I. Ciufolini et al., in {\em General Relativity and John Archibald Wheeler}, edited by I. Ciufolini and R. A. Matzner (Springer, Dordrecht, 2010), p. 371.

\end{thebibliography}

\begin{thebibliography}{999}
\bibitem[{{Demorest} {et al.}(2010)}]{demorestetal10}
 P.~B. {Demorest}, T. {Pennucci}, S.~M. {Ransom}, M.~S.~E. {Roberts}, and
 J.~W.~T. {Hessels},
 Nature, 467, 1081 (2010)
\bibitem[{{Shapiro}(1964)}]{shapiro64}
 I.~I. {Shapiro},
 Phys. Rev. Lett., 13, 789 (1964)
\bibitem[{{Damour} \& {Taylor}(1992)}]{damourtaylor92}
 T. {Damour} and J.~H. {Taylor},
 Phys. Rev. D, 45, 1840 (1992)
\bibitem[{{Lorimer} \& {Kramer}(2005)}]{lorimerkramer05}
 D. {Lorimer} and M. {Kramer},
 Handbook of Pulsar Astronomy
 (Cambridge University Press, Cambridge, 2005)
\bibitem[{{Kramer} {et al.}(2006)}]{krameretal06}
 M. {Kramer}, {et al.},
 Science, 314, 97 (2006)
\bibitem[{{Lattimer} \& {Prakash}(2010)}]{lattimerprakash10}
 J.~M. {Lattimer} and M. {Prakash},
\htmladdnormallink{arXiv:1012.3208}{http://arxiv.org/abs/1012.3208}
\bibitem[{{Lattimer} \& {Prakash}(2001)}]{lattimerprakash01}
 J.~M. {Lattimer} and M. {Prakash},
 Astrophys. J., 550, 426 (2001)
\bibitem[{{Lattimer} \& {Prakash}(2007)}]{lattimerprakash07}
 J.~M. {Lattimer} and M. {Prakash},
 Phys. Rep., 442, 109 (2007)
\bibitem[{{Lattimer} \& {Prakash}(2004)}]{lattimerprakash04}
 J.~M. {Lattimer} and M. {Prakash},
 Science, 304, 536 (2004)
\bibitem[{{Tolman}(1939)}]{tolman39}
 R.~C. {Tolman},
 Phys. Rev., 55, 364 (1939)
\bibitem[{{Oppenheimer} \& {Volkoff}(1939)}]{oppenheimervolkoff39}
 J.~R. {Oppenheimer} and G.~M. {Volkoff},
 Phys. Rev., 55, 374 (1939)
\bibitem[{{Shapiro} \& {Teukolsky}(1983)}]{shapiroteukolsky83}
 S.~L. {Shapiro} and S.~A. {Teukolsky},
 Black Holes, White Dwarfs, and Neutron Stars
 (John Wiley \& Sons, New York, 1983)
\bibitem[{{Rhoades} \& {Ruffini}(1974)}]{rhoadesruffini74}
 C.~E. {Rhoades} and R. {Ruffini},
 Phys. Rev. Lett., 32, 324 (1974)
\bibitem[{{Lindblom}(1984)}]{lindblom84}
 L. {Lindblom},
 Astrophys. J., 278, 364 (1984)
\bibitem[{{Witten}(1984)}]{witten84}
 E. {Witten},
 Phys. Rev. D, 30, 272 (1984)
\bibitem[{{Farhi} \& {Jaffe}(1984)}]{farhijaffe84}
 E. {Farhi} and R.~L. {Jaffe},
 Phys. Rev. D, 30, 2379 (1984)
\bibitem[{{Reynolds} {et al.}(2007)}]{reynoldsetal07}
 M.~T. {Reynolds}, P.~J. {Callanan}, A.~S. {Fruchter}, M.~A.~P. {Torres},
 M.~E. {Beer}, and R.~A. {Gibbons},
 Mon. Not. R. Astron. Soc., 379, 1117 (2007)
\bibitem[{{van Kerkwijk} {et al.}(2011){van Kerkwijk}, {Breton}, \&x
 {Kulkarni}}]{vankerkwijketal11}
 M.~H. {van Kerkwijk}, R.~P. {Breton}, and S.~R. {Kulkarni},
 Astrophys. J., 728, 95 (2011)
\end{thebibliography}
\end{document}